\documentclass[lettersize,journal]{IEEEtran}
\usepackage{amsmath,amsfonts}
\usepackage{algorithmic}
\usepackage{amsmath}
\usepackage{array}
\usepackage{pdfpages}
\usepackage[caption=false,font=normalsize,labelfont=sf,textfont=sf]{subfig}
\usepackage{textcomp}
\usepackage{stfloats}
\usepackage{url}
\usepackage{verbatim}
\usepackage{graphicx}
\usepackage{cite}
\usepackage{hyperref}
\usepackage{pgf}
\usepackage{tikz}
\usepackage{makecell}
\usepackage{bbm}
\usepackage{multirow}
\def\BibTeX{{\rm B\kern-.05em{\sc i\kern-.025em b}\kern-.08em
		T\kern-.1667em\lower.7ex\hbox{E}\kern-.125emX}}
\usepackage{balance}
\usepackage[acronym, section, nonumberlist, nomain]{glossaries}
\newacronym{nn}{NN}{neural network}
\newacronym{iiot}{IIOT}{industrial internet of things}
\newacronym{v2x}{V2X}{vehicle to everything}
\newacronym{ue}{UE}{user equipment}
\newacronym{bs}{BS}{base station}
\newacronym{csi}{CSI}{channel state information}
\newacronym{5g}{5G}{fifth generation}
\newacronym{6g}{6G}{sixth generation}
\newacronym{aoa}{AoA}{angle of arrival}
\newacronym{toa}{ToA}{time of arrival}
\newacronym{rsrp}{RSRP}{reference signal receive power}
\newacronym{ml}{ML}{machine learning}
\newacronym{ai}{AI}{artificial intelligence}
\newacronym{dl}{DL}{deep learning}
\newacronym{cnn}{CNN}{convolutional neural network}
\newacronym{tx}{Tx}{transmitter}
\newacronym{rx}{Rx}{receiver}
\newacronym{de}{DE}{deep ensembles}
\newacronym{mcd}{MCD}{monte carlo dropout}
\newacronym{mtl}{MTL}{multi-task learning}
\newacronym{los}{LOS}{line of sight}
\newacronym{nlos}{NLOS}{non-line of sight}
\newacronym{maml}{MAML}{Model-Agnostic Meta-Learning}
\newacronym{ofdm}{OFDM}{orthogonal frequency division multiplexing}
\newacronym{ura}{URA}{uniform rectangular array}
\newacronym{mse}{MSE}{mean squared error}
\newacronym{crps}{CRPS}{continuous ranked probability score}
\newacronym{cdf}{CDF}{cummulative distribution function}
\begin{document}

\title{Transfer Learning for CSI-based Positioning with Multi-environment Meta-learning}
\author{\IEEEauthorblockN{Anastasios Foliadis\IEEEauthorrefmark{1}\IEEEauthorrefmark{2}, Mario H. Casta\~{n}eda Garcia\IEEEauthorrefmark{1}, Richard A. Stirling-Gallacher\IEEEauthorrefmark{1},  Reiner S. Thom\"a\IEEEauthorrefmark{2}}
	
	\IEEEauthorblockA{\IEEEauthorrefmark{1}\textit{Munich Research Center}, \textit{Huawei Technologies Duesseldorf GmbH}, 
		Munich, Germany \\
		\textit{\IEEEauthorrefmark{2}Electronic Measurements and Signal Processing}, \textit{Technische Universit\"at Ilmenau}, Ilmenau, Germany\\
		\{\href{mailto:anastasios.foliadis@huawei.com}{anastasios.foliadis}, 
		\href{mailto:mario.castaneda@huawei.com}{mario.castaneda}, 
		\href{mailto:richard.sg@huawei.com}{richard.sg}\}@huawei.com, 
		\href{mailto:reiner.thomae@tu-ilmenau.de}{reiner.thomae@tu-ilmenau.de}}}




\maketitle

\begin{abstract}
Utilizing deep learning (DL) techniques for radio-based positioning of user equipment (UE) through channel state information (CSI) fingerprints has demonstrated significant potential. DL models can extract complex characteristics from the CSI fingerprints of a particular environment and accurately predict the position of a UE. Nonetheless, the effectiveness of the DL model trained on CSI fingerprints is highly dependent on the particular training environment, limiting the trained model's applicability across different environments. This paper proposes a novel DL model structure consisting of two parts, where the first part aims at identifying features that are independent from any specific environment, while the second part combines those features in an environment specific way with the goal of positioning. To train such a two-part model, we propose the multi-environment meta-learning (MEML) approach for the first part to facilitate training across various environments, while the second part of the model is trained solely on data from a specific environment. Our findings indicate that employing the MEML approach for initializing the weights of the DL model for a new unseen environment significantly boosts the accuracy of UE positioning in the new target environment as well the reliability of its uncertainty estimation. This method outperforms traditional transfer learning methods, whether direct transfer learning (DTL) between environments or completely training from scratch with data from a new environment. The proposed approach is verified with real measurements for both line-of-sight (LOS) and non-LOS (NLOS) environments. 
\end{abstract}


\section{Introduction}

Accurate user positioning is considered one of the crucial requirements for next generation communication networks \cite{Zhou19,deLima21, Wang22, Chen22}. The ability to localize users is essential for enhancing the efficiency of applications such as \gls*{iiot}. One of the main enablers of high accuracy radio positioning in a wireless network is the support of multiple antennas and larger bandwidths which are made possible by current and future communication networks. 

User positioning in wireless networks has mainly been geometry driven \cite{Kakkavas19}. The \gls*{csi} between a \gls*{ue} and \gls*{bs} is first estimated and from that, some parameters are extracted such as \gls*{toa} and \gls*{aoa}. These parameters reflect in some way the geometry of the environment and therefore can be used to estimate the position of the user. More recently though, there has been an increased focus on using \gls*{dl} to aid or perform positioning by leveraging the ability to collect large amounts of data \cite{Chen17, Yin17, Thomae18, Widmaier19, Butt21}. Positioning methods that are based on DL are able to learn and extract the relevant positioning information from the CSI even in scenario where the classical geometry driven positioning approaches fail \cite{Stahlke2023UncertaintybasedFM}. 

One of the most prominent user positioning methods that leverage the capabilities of DL and \glsplural*{nn} is called fingerprinting \cite{Wang2016CSIPF ,Arnold18, Vieira17}. The idea behind those methods is to train the NN to exploit the information that is embedded in a multi-path channel between the UE and the BS which can be considered a unique fingerprint of the user's location. It is generally divided into two phases. In the offline (or training) phase, a database of CSI fingerprints, measured for different UE positions, along with the respective position is created. This database is used to train a NN to estimate the position. During the deployment phase, the trained NN is used for estimating a UE's position by mapping the CSI of the UE to an estimated position.
 
Positioning methods that are based on CSI fingerprints using DL approaches have shown very promising results achieving sub-meter accuracy in some cases \cite{Foliadis21, Bast20}. Although, despite their advantages against classical methods they are limited in some scenarios. Namely, a NN trained with data from a particular environment fails when employed into some other environment \cite{Stahlke22}. Similarly some change in the environment may cause the model to completely fail because it was trained on data that are not up to date \cite{Foliadis2021ReliableDL, Foliadis2024DeepLB}.

In the context of DL the method of taking a DL-model that was trained on data from some particular distribution, e.g., wireless propagation environment, and applying it in some other distribution is called transfer learning \cite{Goodfellow-et-al-2016}. The paradigm under which the models are designed and trained in such a way to make transfer learning more efficient is called meta-learning \cite{Hospedales2021MetaLearningIN}. In \cite{Owfi23}, a method called \gls*{maml} \cite{Finn2017ModelAgnosticMF} was used to improve generalizability when the training data is limited. The authors of \cite{Owfi23} also propose a variation called task biased MAML (TB-MAML) to further improve transfer learning when the target environment is known. 

In \cite{Foliadis2022MultiEnvironmentBM} a 2-part approach to conventional DL methods of radio-positioning was proposed. This approach is inspired by classical positioning methods where generally a 2-step approach is employed, i.e., environment independent parameter estimation in the first step and environment dependent parameter combination for position estimation in the second step. The main idea presented in \cite{Foliadis2022MultiEnvironmentBM} is to train the first part of the DL model to extract environment independent features of the environment, by training on data from multiple source environments. This way, the environment independent part of the \gls*{dl} model can be reused in a target environment and reduce the amount of data required from that environment (i.e., transfer learning is improved).

In this work we expand upon the work of \cite{Foliadis2022MultiEnvironmentBM} by including uncertainty estimation during initial training of the first part of the DL model as well as during training of the two-part model in target environment. We show that regardless by including the uncertainty estimation during initial training the transferring capabilities of the model are increased. Additionally, in contrast to \cite{Foliadis2022MultiEnvironmentBM} where only \gls*{los} simulated propagation environments were used, we verify our results using real CSI measurements from both \gls*{los} and \gls*{nlos} propagation environments. Lastly, we propose to use gradual unfreezing\cite{Howard2018UniversalLM} during target training in the target environment. Gradual unfreezing is a technique which unfreezes layers of deep neural network models from top to bottom during training. It is used to more efficiently train the model for the target environment and further exploit the information that is embedded in the environment independent first part of the model, especially for small amount of target environment CSI samples.

\section{System Model}

We consider an uplink setup with $N_R$ antennas at a \gls*{bs} and a single antenna at the \gls*{ue}. The UE transmits a reference signal on $N_C$ subcarriers within an \gls*{ofdm} symbol. The received uplink signal is used to estimate the CSI between UE and the BS. The $i$-th estimated channel over the $N_C$ subcarriers between the $k$-th receive antenna and the UE can be described as:

\begin{equation}
	\boldsymbol{h}^k_i = [h_{0, i}^k, h_{1, i}^k, ..., h_{N_C-1, i}^k]^\text{T} \in \mathbb{C}^{N_C}.
\end{equation}
Accordingly, the measured channel between UE and BS's antennas that forms the basis of the CSI fingerprint can be described as:
\begin{equation}
	\tilde{\boldsymbol{H}}_i = [\boldsymbol{h}^0_i, \boldsymbol{h}^1_i, ..., \boldsymbol{h}^{N_R-1}_i] \in \mathbb{C} ^{N_R \times N_C}.
\end{equation}
Eventually, the CSI fingerprint which is used as input to the DL model is determined by stacking the real and imaginary parts of the matrix $\tilde{\boldsymbol{H}}_i$ resulting in a 3-dimensional matrix $\boldsymbol{H}_i \in \mathbb{R} ^ {N_R \times N_C \times 2}$. Subsequently, the CSI fingerprints $\boldsymbol{H}_i$ along with position labels  $\boldsymbol{p}_i=[x_i, y_i] \in \mathbb{R}^2$ are stored and used to train a DL-model for the user positioning task.

We also consider measurements from $N_B$ different environments, with each environment associated with its own dataset $\{(\boldsymbol{H}_{n,i}, \boldsymbol{p}_{n,i})\}_{i=1}^{M_N}$ where $M_n$ are the number of measurements of the $n$-th environment.
Subsequently, the CSI fingerprints of the $n$-th task $\boldsymbol{H}_{n,i}$ along with position labels  $\boldsymbol{p}_{n,i}=[x_{n,i}, y_{n,i}] \in \mathbb{R}^2$ are used to train the parameters $\epsilon_n$ of a DL model $f_{\epsilon_n}(\boldsymbol{H}_{n,i})$ to give an estimate $\tilde{\boldsymbol{p}}_{n,i} = [\tilde{x}_{n,i}, \tilde{y}_{n,i}]\in \mathbb{R}^2$ of the $i$-th position of the user in the $n$-th environment based on the CSI of the user.

Normally, for DL based positioning with CSI fingerprints the DL model $f_n$ is optimized with respect ot the \gls*{mse}:
\begin{equation}
	\mathcal{L}_\text{MSE}(f_{\epsilon_n}) = \mathcal{L}_{x,n} + \mathcal{L}_{y,n},
	\label{eq:mse_loss}
\end{equation}
where $\mathcal{L}_{x,n} = E[|\tilde{x}_{n, i} - x_{n, i}|^2]$, $\mathcal{L}_{y, n} = E[|\tilde{y}_{n, i} - y_{n, i}|^2]$ and $E[\cdot]$ denotes the mean value over the samples in the training set. However, using the MSE criterion does not evaluate the uncertainty of the predictions.

In radio positioning using DL there exist two types of uncertainties. The data dependent uncertainty which is called aleatoric uncertainty reflects the uncertainty that a measurement has about the positioning task and the epistemic uncertainty which reflects the uncertainty of the model\cite{Kendall2017WhatUD}. An example of a measurement with high aleatoric uncertainty is a CSI measurement with low receive SNR. As this type of uncertainty is data-dependent, it can be learned from the available data \cite{Kendall2017WhatUD, Kendall2017MultitaskLU}. In order to estimate it, the loss that has to be minimized is the negative log-likelihood loss (NLL):
\begin{equation}
	\mathcal{L}_\text{NLL}(f_{\epsilon_n}) = \frac{1}{2 (\sigma_{n,i}^x)^2}\mathcal{L}_{x,n}+ \frac{1}{2 (\sigma_{n,i}^y)^2}\mathcal{L}_{y,n} + \log({\sigma_{n,i}^x}{\sigma_{n,i}^y}),
	\label{eq:aleatoric_loss}
\end{equation}

\begin{figure}[b!]
	\centering
	\hspace*{-0.37cm}
	\includegraphics[scale=0.7, trim={50 0 0 0}, clip]{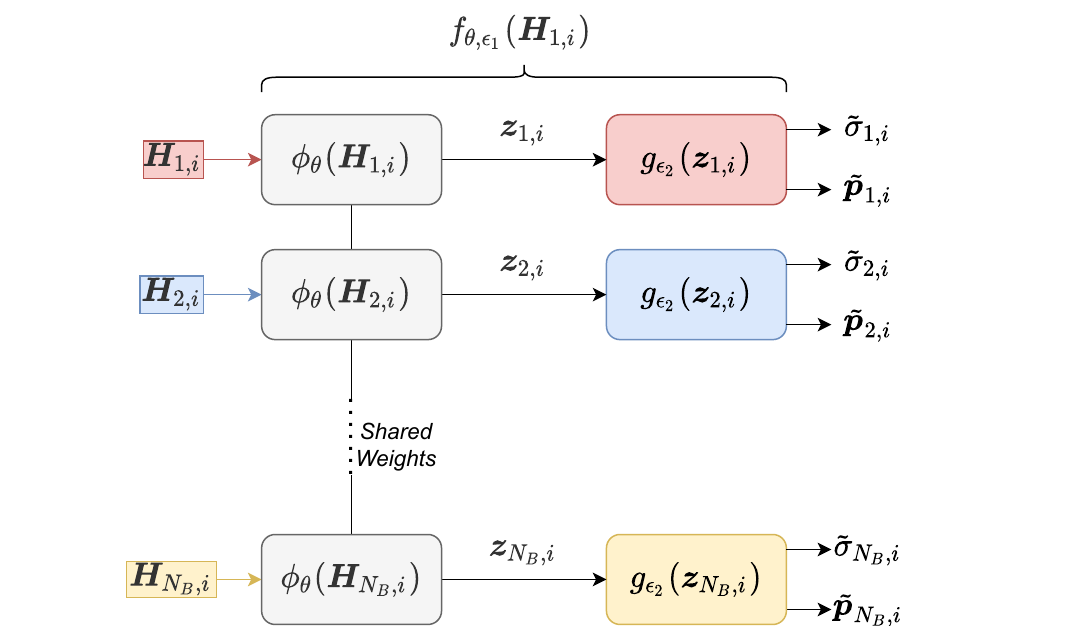}
	\caption{Multi-environment training over $N_B$ environments}
	\label{fig:multi_env_training}
\end{figure}
\begin{figure}[b!]
	\centering
	\hspace*{-0.1cm}
	\includegraphics[scale=0.7, trim={50 0 0 0}, clip]{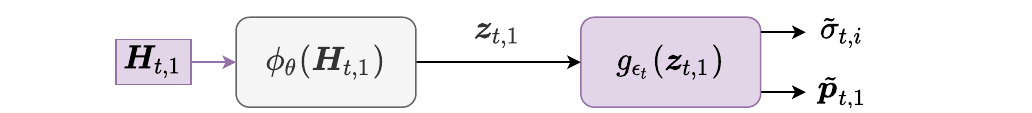}
	\caption{Training of target environment}
	\label{fig:target_model}
\end{figure}
where $\sigma_{n,i}^x$ and $\sigma_{n,i}^y$ are the aleatoric uncertainties for the outputs $x_{n,i}$ and $y_{n,i}$ respectively. When training a model using the NLL loss the output of the neural network has to be modified to include the learned aleatoric uncertainty $\boldsymbol{\sigma}_{n,i} = [\sigma_{n,i}^x, \sigma_{n,i}^y]$. Essentially, the aleatoric uncertainty reflects the inherent uncertainty that each measurement has about the positioning task, e.g., due to noise. This is in contrast to the epistemic uncertainty which reflects the uncertainty of the model itself, e.g., due to low amount of training data. The reader is referred to \cite{Kendall2017WhatUD} for more information about the specific uncertainty types.

We propose the Multi Environment meta-Learning (MEML) approach which is shown in Fig. \ref{fig:multi_env_training}. The basic idea is to consider a two-part structure for the DL model, where the first part is trained with data from all $N_B$ environments while the second part is trained only with data specific to one environment. This way, the first part $\phi_\theta(\cdot)$ is shared across all $N_B$ environments and the second part $g_{\epsilon_n}(\cdot)$ is environment specific and trained on data only from the $n$-th environment. Essentially we define model $f_{\theta, \epsilon_n}(\boldsymbol{H}_{n, i})$ such that $f_{\theta, \epsilon_n}(\boldsymbol{H}_{n, i}) = g_{\epsilon_n}(\phi_\theta(\boldsymbol{H}_{n, i}))$. By employing this type of training, the parameters of the common part $\theta$ are the same regardless of the environment, i.e., they are environment independent while the parameters $\epsilon_n$ depend on the $n$-th environment. The training is carried out by minimizing the sum of the losses for each environment:
\begin{equation}
	\min_{\theta, \epsilon_n} \sum_n\mathcal{L}(f_{\theta, \epsilon_n}).
\end{equation}
When training with the MSE loss, the uncertainty estimate $\boldsymbol{\sigma}_{n, i}$ is not included in the model. It is important to point out that the aim is to make the function $\phi_\theta(\cdot)$ environment independent, i.e., as the function that extracts position relevant parameters from the CSI. Although the function $\phi_\theta(\cdot)$ is common across different environments, the features $\boldsymbol{z}_{n, i}$ are specific to each environment. This is because the distribution of CSI fingerprints $\boldsymbol{H}_{n, i}$, varies with the $n$-th environment. The environment dependent features  $\boldsymbol{z}_n$ are then combined in an environment specific way using the $g_{\epsilon_n}(\cdot)$ function.

Ultimately, the goal of the training of the common part is to acquire an environment independent function which can later be used in a new target environment to aid transfer learning, as shown in Fig \ref{fig:target_model}. The trained part can enable a more efficient minimization of the loss in the new unseen environment $t$ with dataset $\{(\boldsymbol{H}_{t,i}, \boldsymbol{p}_{t,i})\}_{i=1}^{M_t}$:
\begin{equation}
	\min_{\theta, \epsilon_t} \mathcal{L}(f_{\theta, \epsilon_t}),
\end{equation}
where $f_{\theta, \epsilon_t} = g_{\epsilon_t}(\phi_\theta(\boldsymbol{H}_{t,i}))$ and $ g_{\epsilon_t}(\cdot)$ is initialized with random weights before training, while $\phi_\theta(\cdot)$ is initialized with the weights obtained with the MEML training in the source environments.
In our simulations we also consider both MSE loss \eqref{eq:mse_loss} and the negative log likelihood loss \eqref{eq:aleatoric_loss} for the target environment $t$.

Additionally, we propose to use a gradual unfreezing method \cite{Howard2018UniversalLM} of training when transferring the environment independent function to mitigate catastrophic forgetting \cite{Stahlke22}. Since the weights of the function $\phi_\theta(\cdot)$ are already trained, we freeze those layers when transferring to the target environment $t$, therefore only the weights of the function $ g_{\epsilon_t}(\cdot)$ are determined with training in the new environment. After this initial step, when there is no further improvement in the performance of the overall DL-model the weights of the first part are unfrozen and the learning rate is reduced. The training continues again until it reaches its performance limit. This method of training ensures that the first part of the model will preserve the knowledge that it has acquired from the source environments while allowing some adaptation to the new environment. 


For the models which are trained using the NLL loss we propose to use the \gls*{crps} metric \cite{Gneiting2007StrictlyPS} along with the mean error distance between ground truth and estimation. Since these models estimate not only the position but also the uncertainty of the estimation we are interested in a metric that takes into account the reliability of the uncertainty. In other words, a model should be able to identify the cases where the position estimation error is high. The CRPS metric provides a way to quantify the reliability of the uncertainty estimates by comparing the ground truth value to a probability distribution (i.e., the position estimation and it uncertainty). A description of the CRPS metric along with an example that showcases its usefulness is presented in Appendix \ref{app:crps}.

\section{Simulation Setup}

\subsection{Database Description}

To evaluate our proposed approach we use the Dichasus channel measurements described in \cite{dichasus2021}. The measurements were collected with a single-antenna UE transmitting an OFDM signal in the uplink with a bandwidth of 50 MHz and with a pilot sent at every tenth subcarrier out of 1024 subcarriers, i.e., $N_C=103$. The carrier frequency is $f_c = 1.272 \: \text{GHz}$. The ground truth positions are measured with a tachymeter robotic total station, a very precise instrument that tracks the robot’s antenna with a laser with at least centimeter-level accuracy.

\begin{figure}[b]
	\centering
	\hspace*{-0.0cm}
	\scalebox{.99}{\input{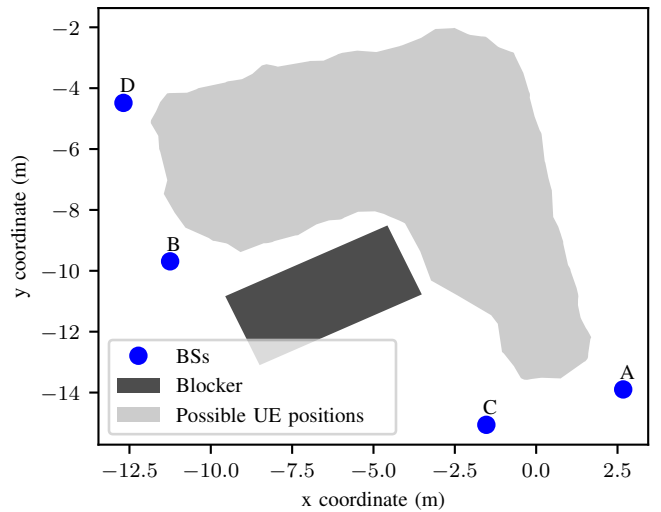}}
	\caption{Layout of considered source environment. DICHASUS Industrial environment \cite{dichasus2021}}
	\label{fig:source_env}
\end{figure}

The source environments are based on data that were collected at four antenna arrays distributed on the corners of an industrial area shown in Fig. \ref{fig:source_env}, where it shown the area where the UE can lie as well as the 4 BS. Each of the antenna arrays consists of a $4 \times 2$ \gls*{ura} with vertical and horizontal antenna spacing of half a wavelength. 
We consider the wireless propagation environment between the UE and each of the BSs as the 4 source environments. Even though this is a single warehouse were all 4 BSs are located, the propagation environment between UE and each of the BSs differs and therefore they can be considered separate environments from the perspective of the BS. This can be verified by training each a model with data from one BS and then testing using data from another BS which results in a huge decrease in performance. Furthermore, it can be seen from the figure that the channels between each possible UE location and every BS include both LOS and NLOS locations.

\begin{figure}[t]
	\centering
	\hspace*{-0.0cm}
	\scalebox{.99}{\input{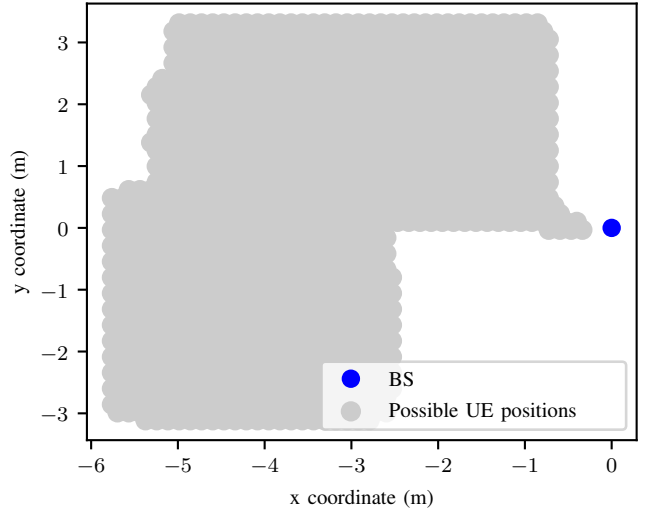}}
	\caption{Layout of considered LOS target environment. DICHASUS Indoor LOS lab room \cite{dichasus2021}}
	\label{fig:target_los}
\end{figure}

\begin{figure}[t]
	\centering
	\hspace*{-0.0cm}
	\includegraphics[scale=0.31]{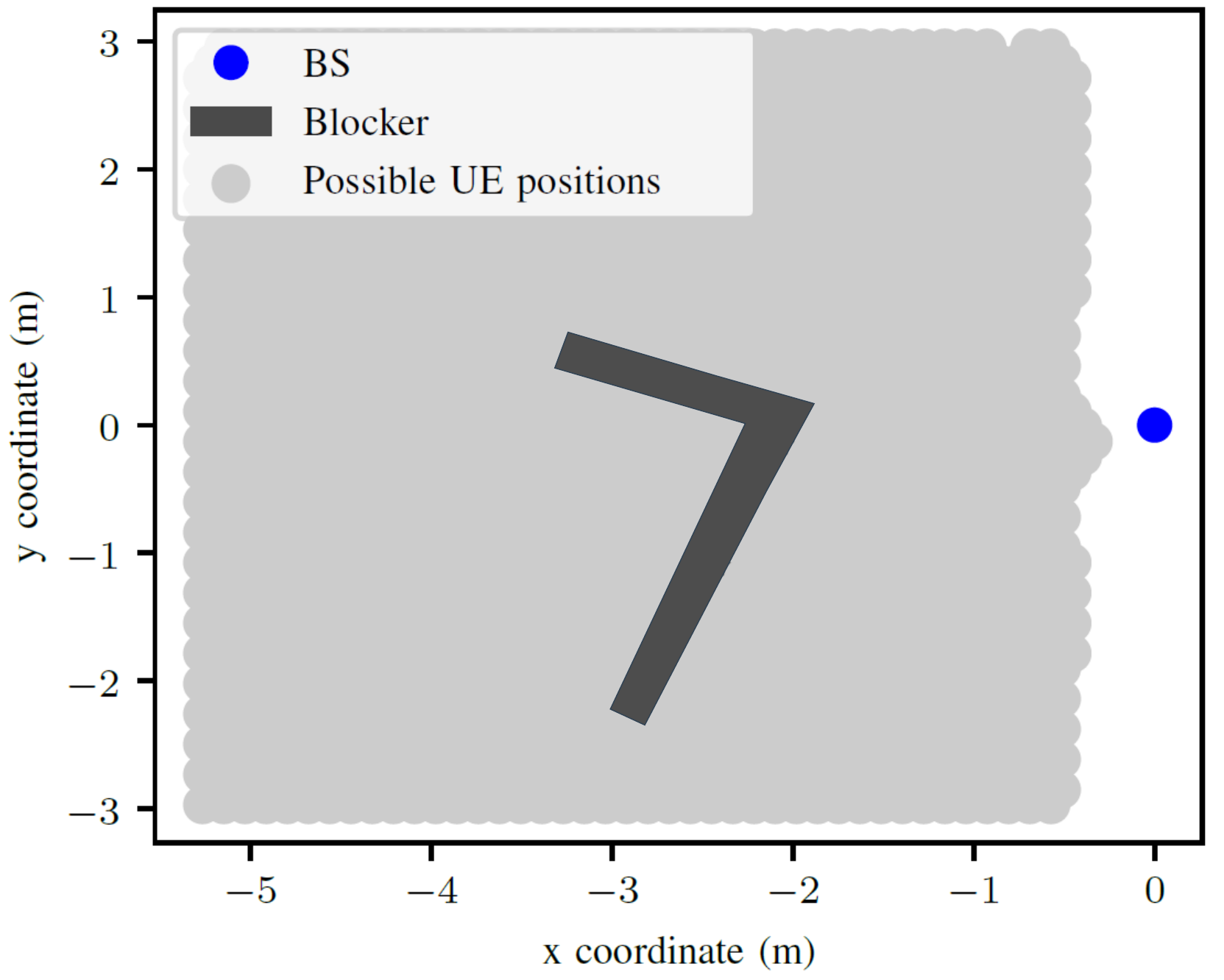}
	\caption{Layout of considered NLOS target environment. DICHASUS Indoor NLOS lab room \cite{dichasus2021}}
	\label{fig:target_nlos}
\end{figure}

For the target environments we consider 2 different datasets based on measurements with one BS in a lab room included in Dichasus measurements \cite{dichasus2021}. One of the datasets includes only LOS measurements while the other includes both LOS and NLOS due to the placement of a blocker which results in a NLOS between some of the UE's positions and the BS. The measurement area for the LOS environment is show in Fig. \ref{fig:target_los} while for the NLOS environment is shown in Fig. \ref{fig:target_nlos}. The measurements for both datasets were collected using one URA with $8 \times 4$ elements at the BS and a single-antenna UE. To align the measurements with those from the source environments, we consider only the measurements from a $4\times2$ subarray within the $8\times4$ array of the BS in the lab room.

\begin{figure}[h]
	\centering
	\includegraphics[scale=0.99]{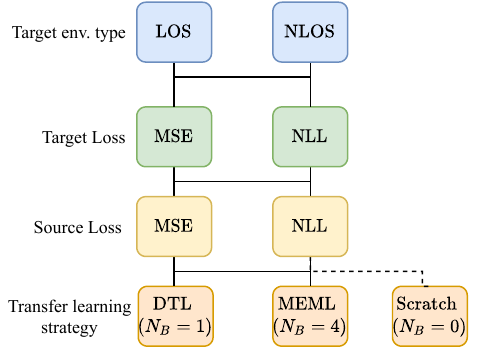}
	\caption{Overview of different simulation options}
	\label{fig:simulation_options}
\end{figure}

In our simulations we compare the different options that we believe would make sense in a real deployment scenario which are shown in Fig. \ref{fig:simulation_options}. Namely, the goal would be to use the trained model and reuse the environment independent part in a target scenario, which in our case are either the LOS or the NLOS target environments. Additionally, depending on the requirements of the deployment one may choose to train the model in the target environment either using the MSE loss or the NLL loss. These options concern with the target environment exclusively therefore we keep them constant and we compare the various training approaches based on source environment settings. Namely we keep constant the target environment (LOS or NLOS) and the target loss (MSE or LLS) and we compare the impact of source loss (MSE or NLL) and the employed transfer learning strategy. The different strategies are the MEML scheme with $N_B=4$ source environments and direct transfer learning (DTL) which involves training a model on one environment and re-training in the target environment. MEML for $N_B=1$ is the same as DTL. Additionally, we include the option of no initial training as baseline, i.e., when training the model from scratch in the target environment.


\subsection{Neural Network and training}

\begin{figure}[h]
	\centering
	\hspace*{-1.0cm}
	\includegraphics[scale=0.65]{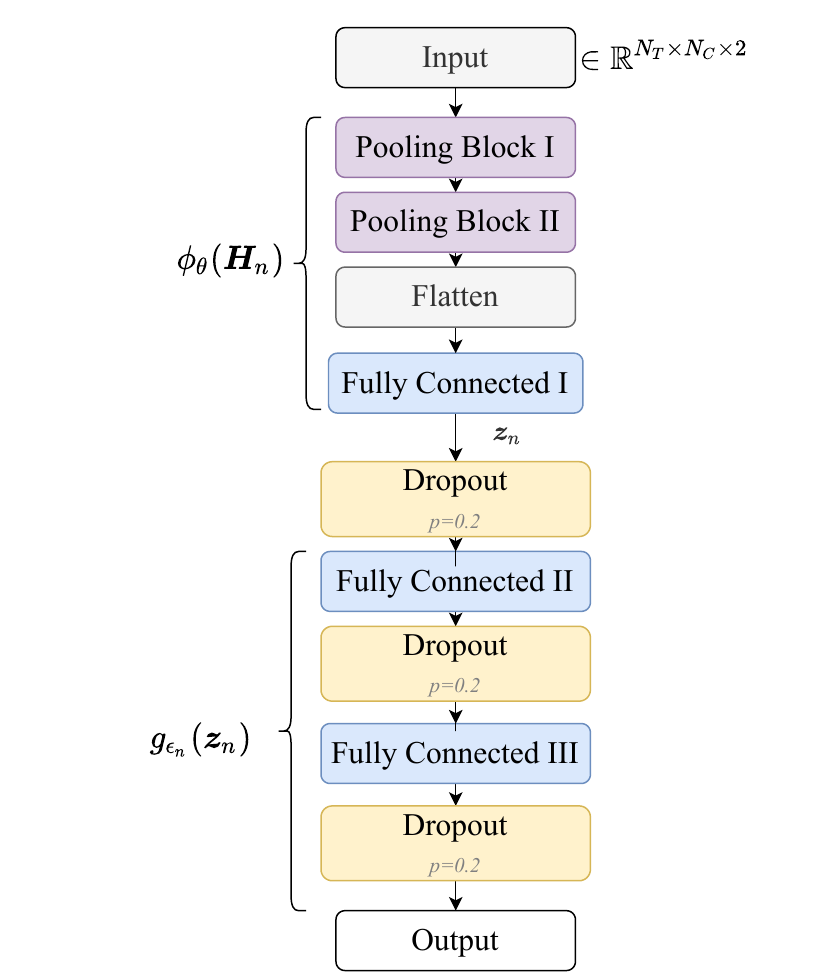}
	\caption{Considered DL model}
	\label{fig:nn_model}
\end{figure}

\begin{figure}[h]
	\centering
	\hspace*{-1.0cm}
	\includegraphics[scale=0.6]{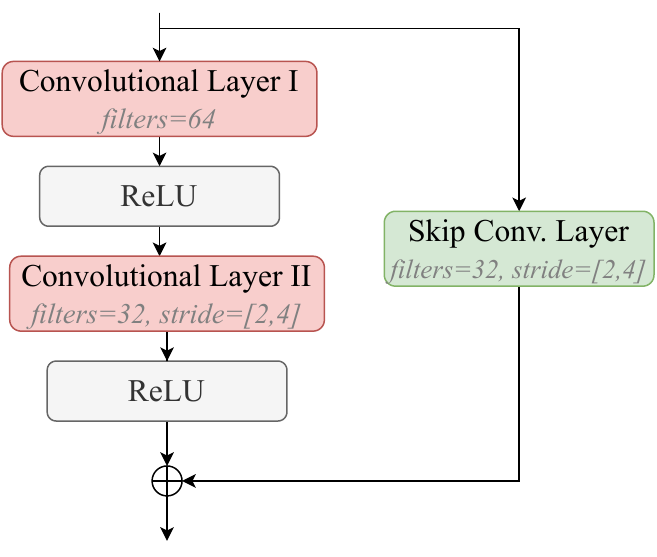}
	\caption{Pooling Block}
	\label{fig:pooling_block}
\end{figure}

For the NN we consider a model that uses convolutional neural networks (CNNs) to form the basis for the pooling block \cite{Zhang2022CSIFingerprintingIL}. The considered DL model is shown in Fig. \ref{fig:nn_model} while the configuration of the considered Pooling Block is shown in Fig. \ref{fig:pooling_block}. The pooling block aims to downsample the input to reduce the computational complexity in the network, while minimizing the information loss, which has also been previously used in the context of AI radio positioning \cite{Zhang2022CSIFingerprintingIL}. Two pooling blocks are placed one after the other and are followed by 3 dense layers with 128 neurons each and finally a dense layer with 2 or 4 neurons for the output, depending on whether the MSE or the NLL loss are used for training. The number of parameters per model are $185\: 922$, with $152 \: 640$ parameters in the first part (environment independent) and $33 \: 282$ in the second part (environment dependent). 

The number of training CSI samples for the source and target environments are shown in Table \ref{tab:samples}. For the source environment the number of samples in the table refers to the number of CSI measurements for each BS. Since we don't consider a deployment scenario in the source environment we don't use any test samples in that environment. For both target environments we incrementally increase the amount of training and validation samples we use from 20 percent to 100 percent of all available samples shown in Table \ref{tab:samples}, while the number of test samples remains constant. 

Additionally, we employ a K-Fold cross-validation method with $k=5$ folds to reduce bias and provide a more accurate estimate of the model's ability to generalize \cite{Goodfellow-et-al-2016}. 
In this approach, the dataset is partitioned into $k$ equally sized, mutually exclusive subsets. The model is then trained $k$ times, each time using $k-1$ as the training set and the remaining fold as the validation set. This process ensures that each data point is used exactly once for validation. The results from each fold are averaged to produce a single estimation. K-fold cross-validation is beneficial in providing a reliable measure of model accuracy and generalizability, as it utilizes all available data for both training and validation.

As an evaluation metric we consider the mean error (ME) of the UE positioning which is given by the euclidean distance between the estimated position and the true position of the UE in the test set. For the models that learn the aleatoric uncertainty through the NLL loss, we use the \gls*{crps} metric \cite{Gneiting2007StrictlyPS}(see Appendix \ref{app:crps}). The results presented in the next section show the average value of the 5 folds.
\setlength{\tabcolsep}{3.8pt}
\renewcommand{\arraystretch}{1.2}
\begin{table}[t!]
	\centering
	\caption{Available Samples for each scenario}
	\begin{tabular}{||c | c | c | c ||} 
		\hline
		\makecell{Samples}&\makecell{Source env.} & \makecell{LOS target env.} & \makecell{NLOS target env.} \\ [0.5ex] 
		\hline\hline
		Training &$60 \: 000$ & $22 \: 234$ & $16 \: 928$\\ 
		Validation & $6000$ & $5558$ & $4234$\\ 
		Testing & N/A & $13 \: 131$ & $5001$\\ 
		\hline
	\end{tabular}
	\label{tab:samples}
\end{table}

\section{Simulation Results}

We test the effectiveness of the MEML approach by learning the function $\phi_\theta(\cdot)$ with data from the source environment shown in Fig. \ref{fig:source_env} and using it to initialize the layers of a new DL models which are trained on data from the LOS and NLOS target environments (Fig. \ref{fig:target_los} and \ref{fig:target_nlos} respectively). We compare the results for different types of source loss (i.e., MSE or NLL). Additionally, we compare the different transfer learning strategies. Namely we compare the MEML approach to direct transfer learning (DTL) whereby a model $f_n$ is trained on one environment only and then retrained on the target environments and also to the case where a model is simply trained from scratch with data from the target environments. We consider those comparisons for each target environment type (LOS or NLOS) and for each target loss type (MSE or NLL) as shown in Fig. \ref{fig:simulation_options}. Lastly, for the best option of each of the target environments and target environments loss types, we compare fine-tuning with gradual unfreezing the model.

\begin{figure}[t] 
	\centering
	\hspace*{-0.5cm}
	\scalebox{0.62}{\input{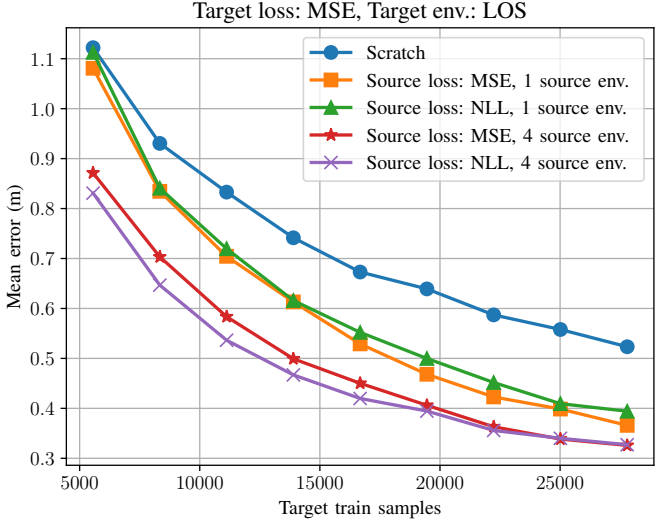}}
	\caption{Mean Error when transferring to LOS env. using the MSE loss}
	\label{fig:mse_LOS}
\end{figure}

Firstly, we transfer the trained $\phi_\theta(\cdot)$ model to the LOS environment using the MSE loss as target loss. The results are shown in Fig. \ref{fig:mse_LOS}. In the figure we see the mean error achieved in the target environment test set for each source training loss and each transfer learning strategy. Regardless of the number of training samples the MEML approach outperforms training the model from scratch or DTL. We also see that there is a slight benefit to training the MEML model using the NLL loss in the source environments, especially for low amount of target training samples. The reason for that is that the model learns better to weigh the important information of each environment during training with data from the source environments.

\begin{figure}[t]
	\centering
	\hspace*{-0.5cm}
	\scalebox{.62}{\input{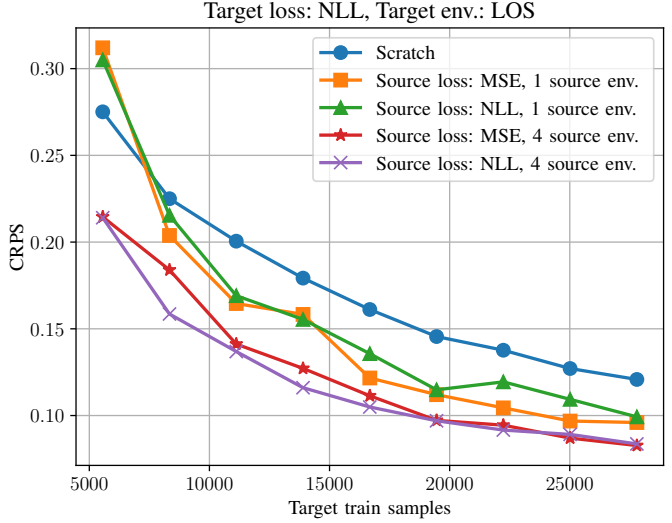}}
	\caption{CRPS when transferring to LOS env. using the NLL loss}
	\label{fig:crps_al_LOS}
\end{figure}

Next, we compare the transfer learning capabilities of different configurations for the LOS target environment when using the NLL loss as target loss function. The results for the mean error in the LOS test set are shown in Fig. \ref{fig:me_al_LOS}. Similarly to when using the MSE loss function as target loss, we see that the MEML method achieves better performance regardless of the target training samples. Since when training using the NLL loss function in the target environment we estimate the uncertainty of each position estimate, in Fig. \ref{fig:crps_al_LOS} we compare the reliability of the uncertainty estimates of different configurations using the CRPS metric. We see that we can reach a similar conclusion, namely that using the MEML not only benefits the position estimate when transferring to a new environment, but also provides a more reliable uncertainty estimation.

\begin{figure}[t]
	\centering
	\hspace*{-0.5cm}
	\scalebox{.62}{\input{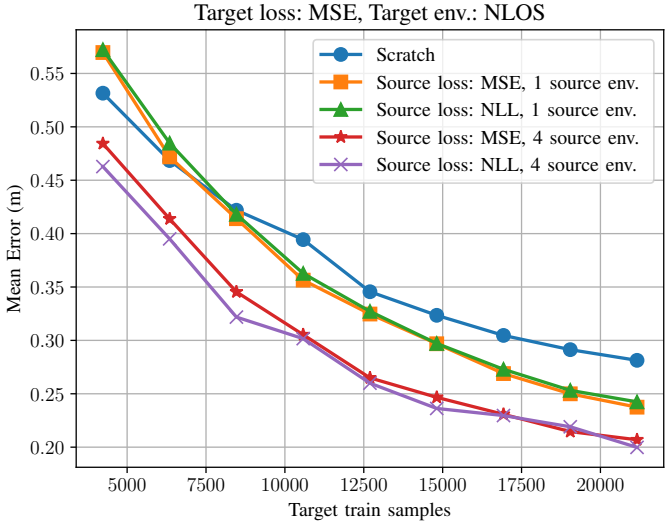}}
	\caption{Mean Error when transferring to NLOS env. using the MSE loss}
	\label{fig:mse_NLOS}
\end{figure}

Now we present the same set of results for the NLOS target environment. In fig. \ref{fig:mse_NLOS} we see the mean error at the test set of the NLOS environment for the different source loss options and transfer learning strategies. In contrast to the LOS target environment we see that DTL may be worse than training the model from scratch in the NLOS target environment when the number of training samples is small. The reason for that is that for small amount of samples the model is not able to adapt to the target environment and leverage the knowledge from the source environments. On the other hand, we again see the benefit of the MEML approach when transferring to a new environment regardless of the number of samples, and again there is a slight benefit to using the NLL loss in the source environment. 

\begin{figure}[t]
	\centering
	\hspace*{-0.5cm}
	\scalebox{.62}{\input{./img/me_al_LOS.pgf}}
	\caption{Mean Error when transferring to LOS env. using the NLL loss}
	\label{fig:me_al_LOS}
\end{figure}

\begin{figure}[t]
	\centering
	\hspace*{-0.5cm}
	\scalebox{.62}{\input{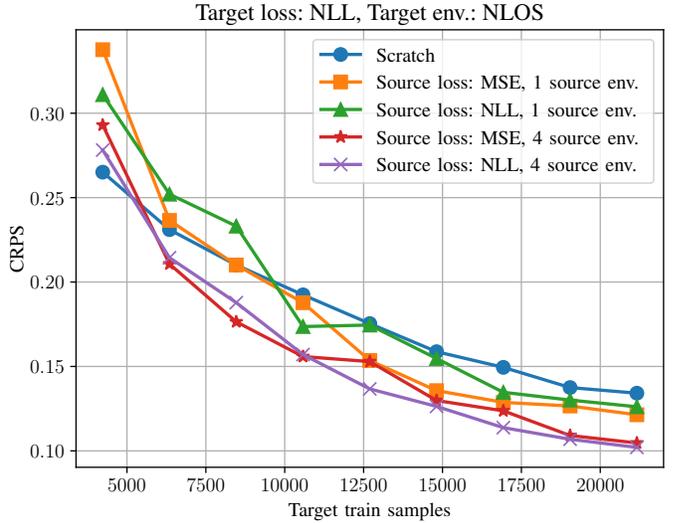}}
	\caption{CRPS when transferring to NLOS env. using the NLL loss}
	\label{fig:crps_al_NLOS}
\end{figure}

In Fig. \ref{fig:me_al_NLOS} and \ref{fig:crps_al_NLOS} we see the mean error and the CRPS metric when transferring to NLOS target environment using the NLL target loss function. It is interesting to see in this case that the MEML method when trained using the NLL source loss shows a clear benefit when transferring with small amount of target training samples. Interestingly, for small amount of training samples the MEML method trained with NLL loss function is the only one that outperforms training the model at the source environment from scratch.

Lastly, we evaluate the gradual unfreezing method of training, and compare it to fine tuning for all MEML configurations.The biggest benefit of gradual unfreezing comes for smaller amount of training samples since with small amount of samples the risk of overfitting is higher. Gradual unfreezing avoids overfitting and catastrophic forgetting of the knowledge that the first part has acquired from the source environments. Therefore we only show the results when using 20\% of the total training samples from each target environment, shown in Table \ref{tab:samples}, i.e., 5558 for the LOS environment and 4232 for the NLOS environment.

\begin{figure}[t]
	\centering
	\hspace*{-0.5cm}
	\scalebox{.62}{\input{./img/me_al_NLOS.pgf}}
	\caption{Mean Error when transferring to NLOS env. using the NLL loss}
	\label{fig:me_al_NLOS}
\end{figure}
In table \ref{tab:mse_LOS} we see the transferring capabilities of the MEML methods for both MSE and NLL source training functions and for fine tuning and gradual unfreezing. We see that the ME when using gradual unfreezing decreases compared to all other transfer learning strategies. Specifically, when using the MEML method trained using the NLL loss function there is a 30 percent decrease in the positioning error compared to just training the model from scratch in the target environment. We can reach similar conclusions when looking the table \ref{tab:al_LOS} which compares the different transfer learning configurations when training using NLL loss function in the LOS target environment. The performance gain for both mean error and CRPS is more than 35 percent when using gradual unfreezing compared to training the model from scratch in the target environment. Another way to look at the benefit of gradual unfreezing is to compare it with the model which is trained from scratch in Figures \ref{fig:mse_LOS}, \ref{fig:me_al_LOS} and \ref{fig:crps_al_LOS}. We see the model which is trained using gradual unfreezing and the MEML transfer strategy needs only around one third of the data to achieve the same performance.

Next, we see the effect of gradual unfreezing when training in the NLOS target environment. In table \ref{tab:mse_NLOS} we see the transferring capabilities of different MEML methods and compare them to training a model from scratch. In contrast to the LOS target environment here the MEML method trained using the MSE source loss function and transferred by gradual unfreezing outperforms the MEML method trained using the NLL source loss function albeit only slightly. The benefit of MEML with gradual unfreezing is 20 percent compared to training a model from scratch. Lastly we show the results when training using the NLL loss function in the target environment. Both ME and CRPS show the benefit of combining MEML with gradual unfreezing achieving more than 25 percent improvement in performance compared to training the model from scratch. 

As mentioned before, gradual unfreezing provides a big benefit for when a small amount of target training samples for both LOS and NLOS target environments. Therefore, we believe it can be beneficial in cases where limited amount of samples is available.

\section{Conclusion}

For positioning with CSI fingerprints, we propose a DL model consisting of two parts, where the first part is trained to be environment independent while the second part of the model depends on the specific environment. The proposed MEML scheme is used to train a DL-model on multiple source environments and the first part of the model, which is environment independent, is used to initialize the weights of a model for a new unseen environment. We investigate the performance of the MEML scheme when it is trained using the MSE loss or the NLL loss and we compare it to the direct transfer learning (DTL) case, i.e., training a model in some environment and fine tuning it in another environment, and also to the case of training a model from scratch in the new environment. We show, that regardless of the number of training samples, the target environment loss function (MSE or NLL) and type of target environment (LOS or NLOS) the MEML method outperforms the other methods. The evaluation is done by means of mean euclidian error between position estimation and ground truth and by use of CRPS metric which takes into account the reliability of the uncertainty estimates of the model. We further propose to use a gradual unfreezing scheme when training the MEML model in the target environment to further improve its positioning accuracy and we achieve a benefit of at least 25 percent in all tested target environment configurations. We also show that by combining gradual unfreezing with the MEML scheme we can reduce the amount of samples used for training in a target environment by one third with no performance reduction.
\begin{table}[t]
	\centering
	\caption{TL capabilities to LOS env. using MSE loss and 5558 Training Samples}
	\begin{tabular}{||c | c | c | c ||} 
		\hline
		\makecell{Source Loss}&\makecell{TL type} & \makecell{Target Training} & \makecell{ME (m)} \\ [0.5ex] 
		\hline\hline
		N/A & Scratch & Fine-tuning & 0.5345 \\ 
		MSE & MEML & Fine-tuning & 0.4292 \\ 
		NLL & MEML & Fine-tuning  & 0.4074 \\ 
		MSE & MEML & Gradual Unfreezing & 0.3994 \\ 
		NLL & MEML &  Gradual Unfreezing  & 0.3796 \\ 
		\hline
	\end{tabular}
	\label{tab:mse_LOS}
\end{table}

\begin{table}[t]
	\centering
	\caption{TL capabilities to LOS env. using NLL loss and 5558 Training Samples}
	\begin{tabular}{||c | c | c | c | c||} 
		\hline
		\makecell{Source Loss}&\makecell{TL type} & \makecell{Target Training} & \makecell{ME (m)} &\makecell{CRPS}   \\[0.5ex] 
		\hline\hline
		N/A & Scratch & Fine-tuning & 0.6349 & 0.278\\ 
		MSE & MEML & Fine-tuning & 0.5074& 0.2627\\ 
		NLL & MEML & Fine-tuning  & 0.4716& 0.2251\\ 
		MSE & MEML & Gradual Unfreezing & 0.4114& 0.1963\\ 
		NLL & MEML &  Gradual Unfreezing  & 0.3925& 0.1769\\ 
		\hline
	\end{tabular}
	\label{tab:al_LOS}
\end{table}

\begin{table}[t]
	\centering
	\caption{TL capabilities to NLOS env. using MSE loss and 4323 Training Samples}
	\begin{tabular}{||c | c | c | c ||} 
		\hline
		\makecell{Source Loss}&\makecell{TL type} & \makecell{Target Training} & \makecell{ME (m)} \\ [0.5ex] 
		\hline\hline
		N/A & Scratch & Fine-tuning & 0.5288 \\ 
		MSE & MEML & Fine-tuning & 0.4803 \\ 
		NLL & MEML & Fine-tuning  & 0.4603 \\ 
		MSE & MEML & Gradual Unfreezing & 0.4168 \\ 
		NLL & MEML &  Gradual Unfreezing  & 0.4194 \\ 
		\hline
	\end{tabular}
	\label{tab:mse_NLOS}
\end{table}

\begin{table}[t]
	\centering
	\caption{TL capabilities to NLOS env. using NLL loss and 4323 Training Samples}
	\begin{tabular}{||c | c | c | c | c||} 
		\hline
		\makecell{Source Loss}&\makecell{TL type} & \makecell{Target Training} & \makecell{ME (m)} &\makecell{CRPS}   \\[0.5ex] 
		\hline\hline
		N/A & Scratch & Fine-tuning & 0.5954 & 0.2752\\ 
		MSE & MEML & Fine-tuning & 0.6595& 0.3795\\ 
		NLL & MEML & Fine-tuning  & 0.5494& 0.2332\\ 
		MSE & MEML & Gradual Unfreezing & 0.4479& 0.2016\\ 
		NLL & MEML &  Gradual Unfreezing  & 0.4364& 0.1996\\ 
		\hline
	\end{tabular}
	\label{tab:al_NLOS}
\end{table}

{\appendix[Continuous ranked probability score (CRPS)]
	\label{app:crps}
	The Continuous Ranked Probability Score (CRPS) is a statistical metric that compares distributional predictions to ground-truth values\cite{Gneiting2007StrictlyPS}. Since the models trained with negative log-likelihood loss predict a variance as well, our intention is to not only achieve high positioning accuracy, but also achieve reliable uncertainty estimations. The estimated uncertainty should be directly the knowledge that the model has about this particular measurement. Since the uncertainty is learned implicitly from the data, there is no ground truth to compare to. Intuitively, if the position estimation error of the model is high, this should be reflected with a corresponding high uncertainty. On the other hand, a low positioning error should correspond to low uncertainty. The \gls*{crps} provides a way to quantify this intuition by comparing a probability distribution (i.e., the output of the model that includes the uncertainty) to a ground truth. It is defined as \cite{Gneiting2007StrictlyPS}:
	
	\begin{equation}
		\text{CRPS}(F, y) = \int (F(x) - \mathbbm{1} _{x\geq y})^2 dx,
		\label{eq:crps}
		\end{equation}
	where $F(x)$ is the \gls*{cdf} of the probability distribution we want to evaluate, $y$ is the ground truth and:
	\begin{equation}
		\mathbbm{1}_{x\geq y} = 
		\begin{cases}
			1  & \text{for  } x\geq y \\
			0 & \text{for  } x < y
		\end{cases}
	\end{equation}
	is an indicator function. It can be shown see that the CRPS is an extension of the mean absolute error but for distributions\cite{Gneiting2007StrictlyPS}. A lower CRPS is desired.
	\begin{table}[b]
		\caption{Toy example ME}
		\centering
		\begin{tabular}{||c c||} 
			\hline
			& \makecell{ME (m)}   \\ [0.5ex] 
			\hline\hline
			Model 1& 0.3117  \\ 
			\hline
			Model 2 & 1.39405  \\ [1ex] 
			\hline
		\end{tabular}
		\label{tab:app_me}
	\end{table}
	\begin{table}[b]
		\caption{Toy example CRPS}
		\centering
		\begin{tabular}{||c c c||} 
			\hline
			& \makecell{Estimated uncertainty  \\$\boldsymbol{\Sigma}_1$} & \makecell{Estimated uncertainty  \\ $\boldsymbol{\Sigma}_2$}  \\ [0.5ex] 
			\hline\hline
			Model 1& 0.13282 & 0.34417  \\ 
			\hline
			Model 2 & 0.80749 & 0.59402 \\ [1ex] 
			\hline
		\end{tabular}
		\label{tab:app_crps}
	\end{table}
	When the model's prediction is a normal distribution $\mathcal{N}(\mu, \sigma^2)$ (as we assume in our case), then Eq. \eqref{eq:crps} is modified to:
		\begin{equation}
		\text{CRPS}(\mu, \sigma^2, y) = \sigma [ \omega (2 \Phi(\omega)-1) + 2 \phi(\omega) - \pi^{-1/2}],
	\end{equation}
	where $\omega = (y - \mu)/ \sigma$.
	
	Next we will present a toy example which illustrates the usefulness of the CRPS metric.
	
	\subsection{CRPS Toy Example}
	We assume that we have build 2 DL-models that give the positions $\boldsymbol{p}_\text{LOS}$ of the $i$-the sample from the LOS environment in Fig. \ref{fig:target_los}. Due to the nature of positioning using fingerprints, each model would have different uncertainty for each position, but for simplicity we assume the the uncertainty of the model is the same regardless of the input sample. These underlying expression for each model which includes the uncertainty is described as:
	\begin{itemize}
		\item Model 1: $\boldsymbol{p}_{1,i} = \boldsymbol{p}_{\text{LOS},i} + \boldsymbol{\eta}_{1}, \, \boldsymbol{\eta}_{1} \sim \mathcal{N}([0, 0], 0.1\boldsymbol{I}_2)$
		\item Model 2: $\boldsymbol{p}_{2,i} = \boldsymbol{p}_{\text{LOS},i} + \boldsymbol{\eta}_{2}, \, \boldsymbol{\eta}_{2} \sim \mathcal{N}([0, 0], 2\boldsymbol{I}_2)$,
	\end{itemize}
	which is the ground truth plus some random noise variable. $\boldsymbol{I}_2$ denotes the identity matrix of dimension $2\times2$. The underlying process that describes model would depend on the amount of samples used during training, the configuration of the model (number of layers and nodes) and other hyperparameters such as learning rate. 
	
	We train each of the models to give an estimate of its own uncertainty, which in reality may differ than the real underlying uncertainty of the model.	
	Intuitively, we prefer the first model to the second one since it is more accurate which is also expressed by the mean error (ME) of each model shown in Table \ref{tab:app_me}. On the other hand, if we have only the second model available, we want the model to accurately estimate its uncertainty, i.e., not to be overconfident, in order to let the application know that the estimates may not be as accurate.  This intuition can be formalized using the CRPS metric.
	
	For the first model let's assume that there are two options for its own uncertainty estimation:
	\begin{itemize}
		\item Low estimated uncertainty: $\boldsymbol{\Sigma}_1 = 0.1\boldsymbol{I}_2$
		\item High estimated uncertainty: $\boldsymbol{\Sigma}_2 = 2\boldsymbol{I}_2$
	\end{itemize}
	If the reported uncertainty of the model is $\boldsymbol{\Sigma}_1$ then we can see that the model accurately estimates its own uncertainty but when the reported uncertainty is $\boldsymbol{\Sigma}_2$ then the model is under-confident. The CRPS metric for the model 1, which is shown in table \ref{tab:app_crps}, increases when the model is underconfident which reflects the decreased performance.
	
	We make the same assumption of the estimated uncertainty for the 2nd model. Namely we assume that the model estimates its own uncertainty as $\boldsymbol{\Sigma}_1$ or $\boldsymbol{\Sigma}_2$. In contrast to the first model, when the second model estimates its uncertainty as $\boldsymbol{\Sigma}_2$ then it is accurate and when the model's estimated uncertainty is $\boldsymbol{\Sigma}_1$ then it means that the model is overconfident. The CRPS metric for the second model with the 2 different uncertainty estimations is shown in Table \ref{tab:app_crps}. 
	
	Out of those 4 cases, shown in Table \ref{tab:app_crps}, we see that the CRPS metric indicates that the first model is better regardless of its estimated uncertainty since it is more accurate, which aligns with our intuition. Additionally, if the first model is under confident it has higher CRPS which is an indication that the model is either underconfident or over confident. The CRPS is decreased when the  uncertainty is estimated accurately . For the second model, the CRPS score is lower when the model is able to accurately estimate its own uncertainty and it increases if the estimated uncertainty diverges from the actual underlying uncertainty of the model. 
	
	In this Appendix we showed the usefulness of using the CRPS metric for positioning when we want to estimate the reliability of uncertainty estimates in addition to positioning accuracy.

		}
	
\bibliographystyle{IEEEtran}
\bibliography{IEEEabrv,../Dissertation/literatur.bib}
%
%
%
%

\vfill

\end{document}